# THEY DON'T LOOK DIFFERENT, BUT THEY'RE NOT THE SAME: FORMAL RESEMBLANCE AND INTERPRETIVE DISPARITY IN THE CONSTRUCTION OF TEMPORAL FREQUENCY DISTRIBUTIONS

William A. Brown

Department of Anthropology
and
Center for Social Science Computation and Research (CSSCR)
University of Washington

brownw@uw.edu

**Abstract**
In archaeological and paleontological demographic temporal frequency analysis (dTFA), a handful of protocols for generating temporal frequency distributions (*tfd*s) have emerged based on the aggregation not of single-point timestamps but instead of constituent temporal distributions, including probability summation, kernel density estimation, and human occupation index calculation. While these protocols bear a striking algebraic resemblance to one another, they are motivated by the desire to contain fundamentally different sources of uncertainty, leading to detailed differences in procedure as well as fundamental differences in the interpretation of the resulting *tfd*. Rather than assuming that one technique can fulfil dual purposes based on its formal resemblance with another, the joint containment of multiple sources of uncertainty therefore warrants the adoption of propagation-of-uncertainty techniques in *tfd* construction.

## 1. Introduction

The definitive empirical feature of demographic temporal frequency analysis (dTFA) is the temporal distribution (or 'temporal frequency distribution,' abbreviated *tfd*) of recovered archaeological or paleontological materials. In the possession of such distributions, dTFA is predicated on the proposition that these may be treated as a proxy census time series, providing information on the temporal dynamics of regional population size and growth as these have varied over time, under certain well-controlled research conditions (Rick, 1987; Chamberlain, 2006: 131-132; Drennan et al., 2015: 12-14; cf. Brown, 2017: Supplementary Information for an extensive list of publications including *tfd*s, most of which are tacitly if not explicitly interpreted as proxy census records).

Over its nearly five decades of existence (taking Haynes' 1969 paper as a provisional starting point), dTFA has seen the application of several different procedures for aggregating *tfd*s, though the histogram (including the histogram presenting as a polygon) dominated the first two and a half decades of the program's existence. However, the probabilistic nature of most chronometric methods employed by archaeologists and paleontologists presents a nontrivial challenge for histogram aggregation, in the form of potential bin mis-assignment. Of the various measures prescribed and taken to mitigate this problem, the ascendant strategy has been the application of an alternative method for *tfd* aggregation – probability summation – introduced in the 1970s (Black and Green, 1977, cited by Dye and Komori, 1992; sources cited in Weninger, 1986: 21) though gradually gaining in popularity in Anglophone archaeology and paleontology only after the late 1980s (e.g., Anderson, 1989; Dye and Komori, 1992; Shott, 1992; Batt and Pollard, 1996).

More recently, kernel density estimation has emerged in dTFA as yet another method for *tfd* aggregation (Louderback et al., 2010; Grove, 2011; Tallavaara et al., 2014; Tallavaara, 2015; Baxter and Cool, 2016; Fitzhugh et al., 2016; Weitzel and Codding, 2016; Brown, 2017), drawing on non-parametric inferential statistics. No later than this new approach's arrival into dTFA, the algebraic likeness between probability summation and kernel density estimation was noted (Louderback et al., 2010; Grove, 2011; Baxter and Cool, 2016): both protocols involve the aggregation not of single-point timestamps but instead multiple temporal distributions (hereafter, 'constituent distributions') into *tfd*s, specifically through summation across such constituent distributions.

To be more formally explicit, let an individual constituent distribution be denoted with the index $i$, from 1 to sample size $n$. Further let the $i$th constituent distribution comprise a series of non-negative density functions along the timeline, generically denoted $\varphi_i(t)$.[1] In the standard cases both of probability summation and of kernel density estimation, these distributions are 'equally massed,' i.e. the total mass under any one constituent distribution equals the total mass of any other:[2]

$$\int_{\infty}^{-\infty} \varphi_1(t)\, dt = \cdots = \int_{\infty}^{-\infty} \varphi_n(t)\, dt \quad (1).$$

---

[1] In the present paper, *t* is understood as a single point in time along the 'calendric years before present' or 'cal BP' timeline, such that *t* decreases as time elapses. By convention, the origin of the cal BP timeline is 00:00 on 1 January 1950 CE.

[2] Note that because the cal BP timeline decreases as time elapses, the lower and upper boundaries of the integral are reversed. In theory, the lower boundary, ∞, refers to 'the beginning of time' and the upper boundary, -∞, to 'the end of time.' In practice, however, the timeline is truncated at the lower end either by the beginning of the archaeological or paleontological record or by 50,000 cal BP (approximately the limit of $^{14}$C dating), and at the upper end either at a given species' extinction or at the moving present (which at the moment is 14:56 on 1 August 2017).

Furthermore, for convenience if not by necessity, these equally massed distributions are often set to integrate to unity, which may require the application of a normalizing constant (denoted $C_N$) provided that such normalization is not already intrinsic to the density function:

$$1 = C_{N,1} \times \int_{\infty}^{-\infty} \varphi_1(t)\, dt = \cdots = C_{N,n} \times \int_{\infty}^{-\infty} \varphi_n(t)\, dt \quad (2a)$$

$$= \int_{\infty}^{-\infty} C_{N,1}\, \varphi_1(t)\, dt = \cdots = \int_{\infty}^{-\infty} C_{N,n}\, \varphi_n(t)\, dt \quad (2b).$$

The final step in both protocols is to aggregate their constituent distributions through time-dependent summation across all $n$ constituent density functions, having the general form:

$$\omega(t) = C \times \sum_{i=1}^{n} \varphi_i(t) \quad (3),$$

where $\omega(t)$ is a generic *tfd* function and $C$ is an optional scaling constant (see Fig. 1). When this scaling factor is set to 1 and all constituent distributions integrate to unity, the *tfd* function integrates to sample size. Conversely, when the scaling factor is set to the reciprocal of sample size and all constituent distributions integrate to unity, the *tfd* function –

$$\omega(t) = \frac{1}{n} \times \sum_{i=1}^{n} \varphi_i(t) \quad (4)$$

– likewise integrates to unity (qualifying this scaling factor as a normalizing constant). Note that the expression in Eq. 4 is nothing other than an unweighted average of the function $\varphi_i(t)$ across all constituent distributions at time $t$. Fig. 1 provides a graphic illustration of this summation operation.

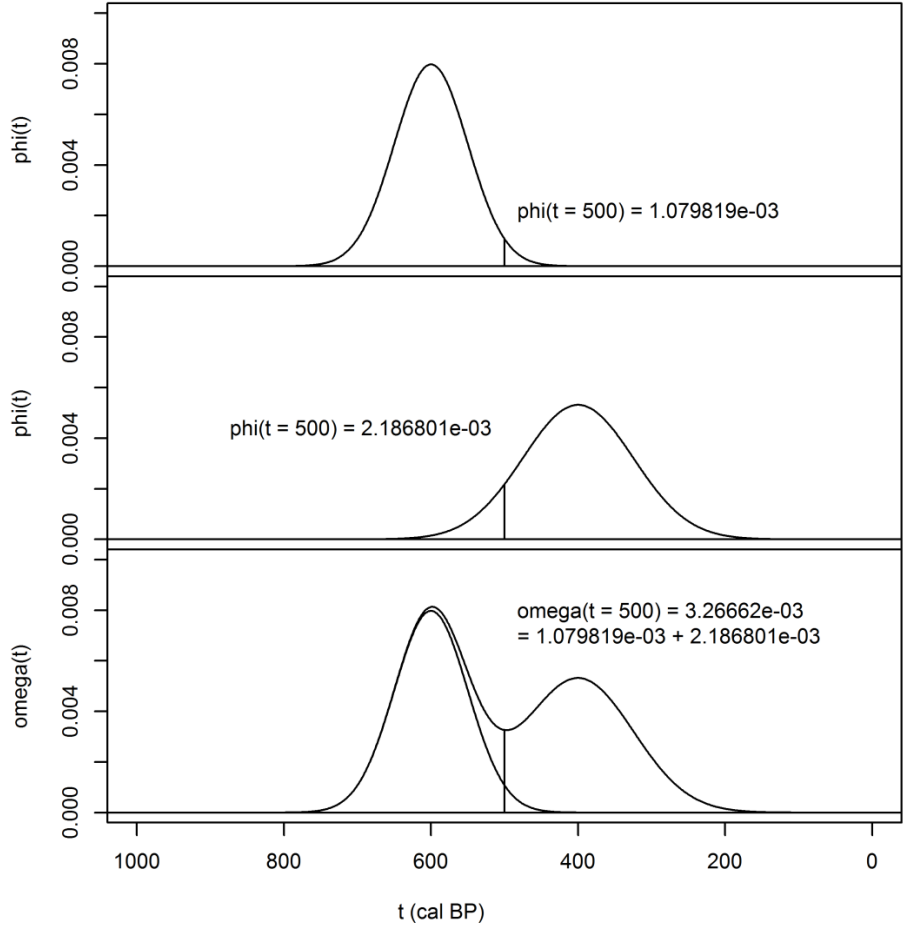

**Fig. 1.** Upper panel: a Gaussian constituent distribution with a mean of 600 cal BP and a standard deviation of 50 years (area under the curve = 1). The vertical line shows the distribution's density at 500 cal BP, $\varphi(t = 500)$. Middle panel: a second Gaussian constituent distribution, with a mean of 400 cal BP and a standard deviation of 75 years (area under the curve = 1). Once again, the vertical line shows the distribution's density at 500 cal BP. Lower panel: a *tfd* generated by summing across the two constituent distributions shown in the upper and middle panels (area under the curve = 2). The visual effect is one of stacking or draping one constituent distribution upon the other. The vertical line shows the distribution's summed density at 500 cal BP, $\omega(t = 500)$, equaling the sum across both constituent distributions' $\varphi_i(t = 500)$ functions.

Algebraic resemblance notwithstanding, the probability summation approach is not a logical extension of kernel density estimation (contra Louderback et al., 2010: 368), nor is the converse true. By implication, while it may be tempting to assume that the application of one obviates the need or desirability to apply the other – a proposition tacitly accepted by Grove (2011: 1017-1018) and Tallavaara (2015: 26) – we would be ill-advised to proceed in this manner. Historical consideration of each protocol's motivation, as well as the formal characteristics and interpretations of their respective constituent distributions, warns against both their conceptual and operational conflation.

This paper systematically summarizes the historical origins of and detailed differences between these two and other formally similar methods for *tfd* aggregation, demonstrating fundamental differences in their theoretical content. The second and third sections of this paper unpack these two protocols further, discussing the inferential problem motivating each; the interpretation, formal characteristics, and idiosyncratic behaviors of their respective constituent distributions; and the interpretation of their respective aggregate *tfd*s. The fourth section then reviews two alternative approaches for the joint mitigation of the two problems addressed by these methods – composite kernel density estimation introduced by Brown (2017), and weighted kernel density estimation having its formal origins in Weninger (1986: 32-33) but reinterpreted by the present author (Brown, 2015: 143; Fitzhugh et al., 2016: Supplementary Material). Differing interpretations of this latter protocol are further discussed in the fifth section. Another *tfd*-generating protocol, introduced by Maschner and colleagues (2009) and resembling the composite kernel density estimation technique in form but not motivation, is discussed in the sixth section.

**2. Chronometric error and probability summation**
*2.1. Motivation*

As already noted, probability summation emerged in dTFA in response to the growing awareness that the probabilistic character of chronometric estimates based on $^{14}$C and similar dating methods presents a special challenge for histogram bin assignment. Imagine a scenario in which the *tfd* is expressed as a histogram whose bins move backward in time from 0 cal BP at 200-year intervals. Further imagine that the age of an event that transpired in 1115 cal BP is estimated at 1000±100 cal BP (following a normal distribution). If the central estimate, 1000 cal BP, is used as the basis for bin assignment, this will result in a mis-assignment, in this case to the consequent interval $[1000, 800)$ rather than the correct interval $[1200, 1000)$. As a result, the true bin count is deflated while the adjacent bin count is inflated to the same degree (a "coverage error" in the parlance of formal demography; Preston et al., 2001: 211).

One strategy for mitigating this problem, still in use particularly in exploratory dTFA contexts, is the adoption of wider bins, resulting in the assignment of a greater fraction of data points to their proper bins. This procedure is relatively efficient in that it does not require any special effort to formally incorporate information about measurement error into the *tfd* aggregation protocol; a wide bin width is simply selected, usually equaling two to four times the typical standard error for a given data set (if not more). This strategy is sometimes further enhanced by removing those age estimates having particularly imprecise errors from the sample. Even so, this strategy is problematic on several counts (cf. Dye and Komori, 1992: 36):

- Working with wide bins means accepting a coarser analytical grain than might otherwise be warranted by sampling error considerations alone (Freedman and Diaconis, 1981; Silverman, 1986: 7-11).
- Culling age estimates with large errors inflicts a reduction in sample size not motivated by concerns for accuracy of the data thus removed.

- While the strategy is expected to reduce the fraction of mis-assigned data points, some will nevertheless lie near enough to the boundaries of their true bins that some will still be mis-assigned to adjacent bins.

The heavy reliance both of archaeologists and of latest quaternary paleontologists on $^{14}$C dating further exacerbates the problem of working with histogram *tfd*s, both because the precision of uncalibrated $^{14}$C age estimates can be a poor guide to the precision of their calibrated age estimates and because their calibrated probability distributions are often highly irregular (multimodal and/or asymmetric), making the identification of a reliable single-point age estimate problematic (Telford et al., 2004; Bronk Ramsey, 2009a: 353-354; cf. Hoff, 2009: 21-22; Gelman et al., 2013: 33-34).

Consequently, the desire for a *tfd* aggregation method that better accommodates chronometric uncertainty led to the introduction of the probability summation method (Weninger, 1986: 21, noting however that Weninger employs the term 'histogram' to refer to *tfd*s generated through probability summation; Dye and Komori, 1992: 36).

*2.2. Constituent distribution*

The probability summation protocol's constituent distribution is a probability distribution quantifying our varying degree of belief in the true value of the *i*th data point's timestamp over the timeline. In the special but archaeologically pervasive case of $^{14}$C-based timestamp estimation, this is usually a Bayesian posterior probability distribution, $p(\tau_i = t | r_i, s_i, m_i)$, where $r_i$ and $s_i$ are the $^{14}$C assay and corresponding measurement error informing the estimate and $m_i$ is the calibration curve or forward map model used to calculate $r_i$ and $s_i$ for this data point. As is true of probability density functions generally, this function is non-negative and integrates to unity over the timeline. Probabilistic age estimates based on other chronometric techniques may also serve as building blocks for probability summation, though few if any examples of such have actually been employed in archaeological case studies (but cf. Contreras and Meadows, 2014, whose simulations based on OxCal's 'C_Simulate' function generate samples of generic, normally distributed chronometric estimates).

*2.3. tfd nomenclature and generative protocol*

While *tfd*s generated through probability summation have gone under many names over their >2 decades of application – e.g., "smooth frequency curves" (Black and Green, 1977) or "cumulative probability curves" (Anderson, 1989) if aggregated from uncalibrated $^{14}$C age estimates (Dye and Komori, 1992; cf. Louderback et al., 2010) or "annual frequency distribution diagrams" if aggregated from calibrated ones (Dye and Komori, 1992); "cumulative-error spectra" (Shott, 1992: Fig. 5, citing an unpublished manuscript by S.W. Robinson) – the term that has stuck is the 'summed probability distribution' (*spd*), whose temporal density function will be formally denoted $\omega_{spd}(t)$ in the following.

In the case of $^{14}$C-based *spd* aggregation, the protocol is complete with the summation across all $n$ posterior probability density functions at any given point along the timeline, with optional rescaling by a constant scaling factor $C$:

$$\omega_{spd}(t) = C \times \sum_{i=1}^{n} p(\tau_i = t | r_i, s_i, m_i) \quad (5)$$

(compare Eq. 3). Fig. 2 graphically illustrates the output of this procedure for a two-observation sample, whose two constituent posterior distributions are calibrated using a linear interpolation of the IntCal13 model (Reimer et al., 2013), assuming a uniform prior distribution (Bronk Ramsey, 2009a:342), from the two $^{14}$C ages 1760±25 BP and 1620±40 BP (compare Fig. 1; cf. Contreras and Meadows, 2014: Fig. 1; Bettinger, 2016: Fig. 1B).

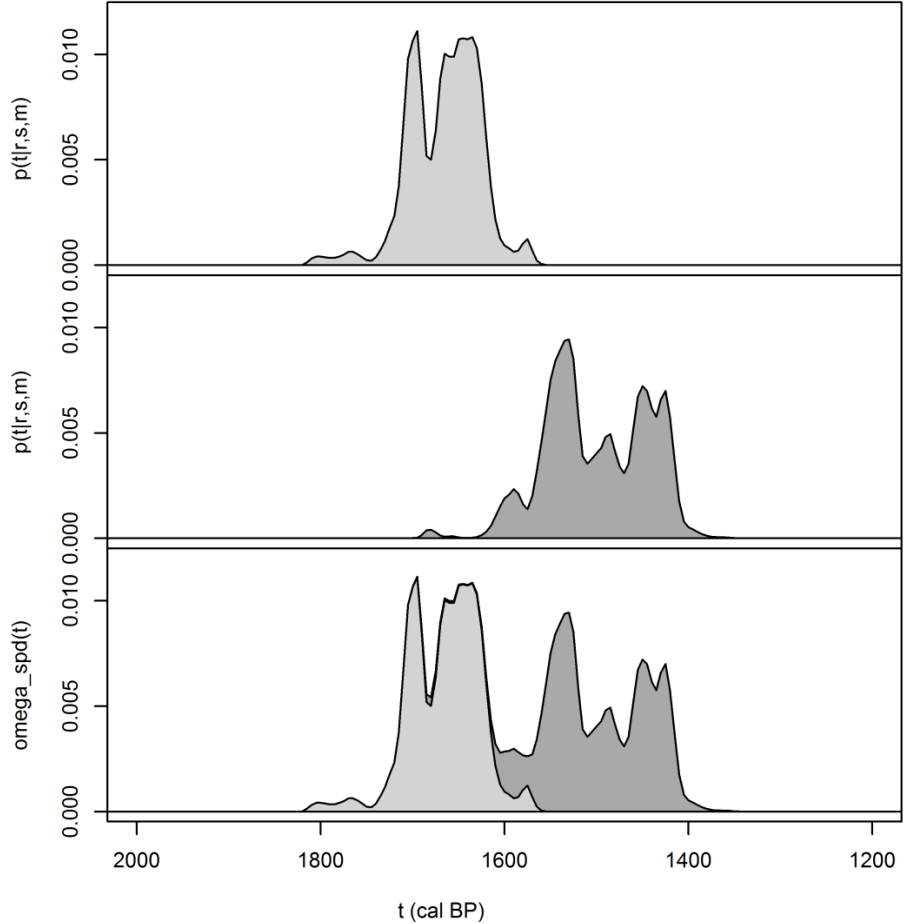

**Fig. 2.** Upper panel: posterior probability distribution over the timeline calibrated using a linear interpolation of the IntCal13 model and assuming a uniform prior distribution from 1760±25 BP. Middle panel: posterior probability distribution over the timeline calibrated using a linear interpolation of the IntCal13 model and assuming a uniform prior distribution from 1620±40 BP. Lower panel: a *spd* generated from these two posterior distributions. As in the generic example given in Fig. 1, the visual effect here is one in which the *spd* comprises a set of stacked posterior distributions. In this case, the area under the *spd* curve is 2 (i.e., the scaling constant in Eq. 5 equals 1, simplifying out of the equation).

*2.4. Interpretation*

Given the uncertainty that defines probabilistic age estimation, the *spd* embodies a "good approximation to the true frequency distribution of events" (Bronk Ramsey, 2001:361) or "best estimate for the chronological distribution of the *items dated*" (Bronk Ramsey, 2005; emphasis added to accentuate the status of the *spd* as a sample distribution). While in the OxCal v. 3 manual Bronk Ramsey (2005) asserted that this procedure is "difficult to justify statistically," such justification is in fact attainable, if somewhat technically complicated. Such a justification is offered in the remainder of this subsection.

*tfd*s aggregating exact timestamps are relatively straightforward to generate, provided such perfect knowledge is available or assumed. Conversely, when chronometric exactitude is unobtainable, the next best approach is to attempt a probabilistic containment of uncertainty surrounding the $n$ unknown timestamps characterizing the sample. This entails a parameter estimation problem in an $n$-dimensional parameter space, denoted $T$, containing all possible timestamp combinations that may characterize the sample. Let **t** denote a single location within this parameter space, abbreviating a vector whose elements constitute a series of possible true timestamps for the data points included in the sample:

$$\{\pmb{\tau} = \mathbf{t}\} = \{\tau_1 = t_1, \dots, \tau_n = t_n\} \in T \quad (6).$$

Note that while any single location in the parameter space unequivocally implies an exact-point *tfd*, the inverse does not hold; that is, when $n \geq 2$, any given exact-point *tfd* is equally implied by two or more separate locations in the parameter space. For example, all three locations in an $n = 3$ parameter space –

$$\{\tau_1 = 100, \tau_2 = 100, \tau_3 = 50\} \quad (7a)$$

$$\{\tau_1 = 100, \tau_2 = 50, \tau_3 = 100\} \quad (7b)$$

$$\{\tau_1 = 50, \tau_2 = 100, \tau_3 = 100\} \quad (7c)$$

– imply exactly the same exact-point *tfd*:

$$\{\omega(100) = 2, \omega(50) = 1\} \quad (8).$$

The Bayesian containment of uncertainty across any such parameter space involves first of all the specification of a joint prior probability distribution describing varying degrees of belief between different locations in the parameter space prior to the provision of chronometric information,

$$p(\mathbf{\tau} = \mathbf{t}) = p(\tau_1 = t_1, \ldots, \tau_n = t_n) \quad (9),$$

and second of all the updating of this joint distribution in light of chronometric information, expressed as a joint posterior probability distribution

$$p(\mathbf{\tau} = \mathbf{t}|\mathbf{r}, \mathbf{s}, \mathbf{m}) = p(\tau_1 = t_1, \ldots, \tau_n = t_n | r_1, \ldots, r_n, s_1, \ldots, s_n, m_1, \ldots, m_n) \quad (10).$$

Fig. 3 describes the joint posterior probability distribution implied by the two posterior distributions presented in Fig. 2. Assuming that these two age estimates are statistically independent of one another, the joint probability is simply the product of the two marginal distributions

$$p(\tau_1 = t_1, \tau_2 = t_2 | r_1, s_1, m_1, r_2, s_2, m_2) = p(\tau_1 = t_1 | r_1, s_1, m_1) \times p(\tau_2 = t_2 | r_2, s_2, m_2) \quad (11).$$

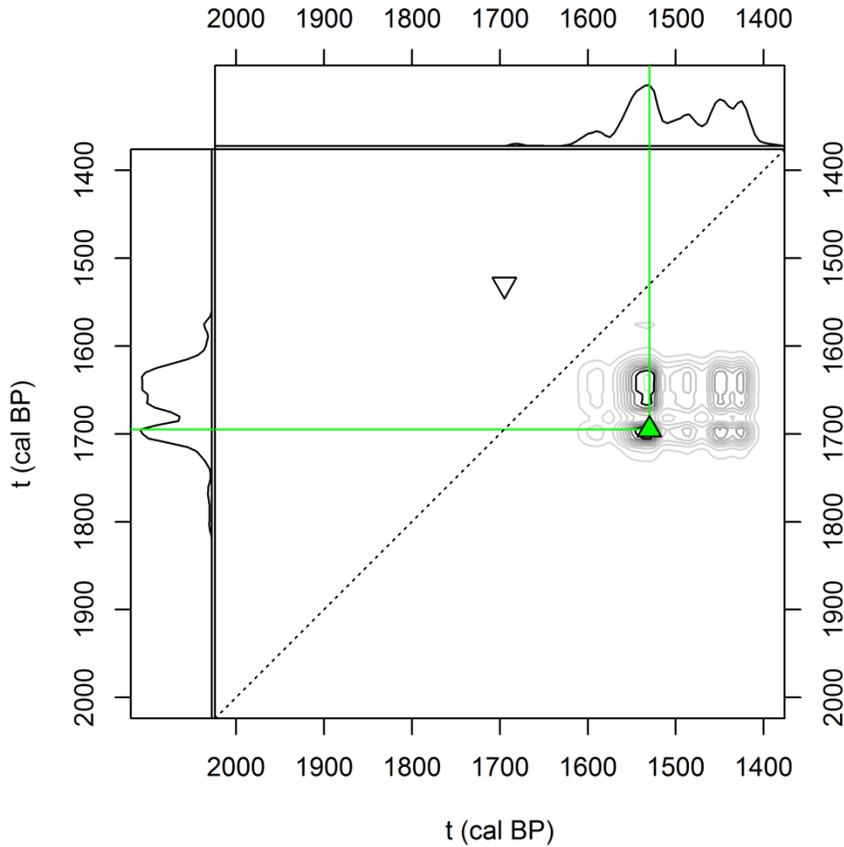

**Fig. 3.** Joint posterior probability distribution over a two-dimensional parameter (timestamp) space (contour plot) and its two marginal distributions (left and upper panels). The left and upper marginal distributions are identical to the upper and middle distributions shown in Fig. 1, respectively, and are shown at identical scales for comparison with each other. Contour lines in the joint probability plot darken as joint probability increases, this surface being the product of the two marginal distributions per Eq. 11. The MAP for the joint distribution (Eq. 12) lies at the intersection of the two marginal distributions' MAPs, (1530, 1695), denoted with the green/shaded triangle and , implying the *tfd* $\{\omega(1530) = 1, \omega(1695) = 1\}$. In the context of the hypothetical MC simulation described in the text, which samples across this joint distribution (Eq. 13), this MAP is the single most likely out of all possible guesses. However, note that a second pair of coordinates, (1695, 1530), located in a low-probability region of the parameter space (inverted triangle) and therefore less likely to be randomly drawn, implies the same *tfd* as the MAP. The total probability of drawing this *tfd* equals the sum of the joint probabilities at these two distinct locations. The diagonal dotted line transecting the parameter space contains all cases where the true timestamp of both data points is identical (such that integrating the joint probability function along this line would give the total posterior probability that the two data points share the same age). Any line drawn perpendicular to this main diagonal, with a symmetrical span around it, will identify another pair of separate locations in the parameter space that imply the same *tfd*.

The single best guess at the true temporal distribution of the sample would involve maximizing this joint posterior function across the parameter space:

$$\hat{\mathbf{t}}_{MAP} = \arg\max_{T} p(\mathbf{\tau} = \mathbf{t}|\mathbf{r}, \mathbf{s}, \mathbf{m}) \quad (12),$$

where $\hat{\mathbf{t}}_{MAP}$ abbreviates the maximum *a posteriori* (MAP) or posterior mode estimator over this space (Hoff, 2009: 21; Murphy, 2012: 4; Gelman et al., 2013: 33, 313-318). However, as in any case of posterior inference, single best guesses are far from sufficient when taken alone. Usually, the recommended approach in posterior inference is to supplement the point estimator with a credible interval or region (CI or CR) containing uncertainty around it (Hoff, 2009: 41-43; Gelman et al., 2013: 33-34; Telford et al., 2004; Bronk Ramsey, 2009a: 353-354). However, for multi-parameter estimation problems, the feasibility of efficiently reporting CRs is low and their interpretability lower still. In addition, since we are more directly interested in estimating the sample *tfd* than the location of the sample in an $n$-dimensional timestamp parameter space, and since any given sample *tfd* has a one-to-many relationship with the locations constituting the parameter space (as exemplified by Eqs. 7-8 or by the two locations in Fig. 3 identified by triangles), an alternative approach to posterior inference is desirable.

     One such approach is model-averaging (sources cited in Hoff, 2009: 170 and Gelman et al., 2013: 193; cf. Bronk Ramsey 2009b). In the context of sample *tfd* estimation, model-averaging involves the calculation of a weighted average of all possible *tfd*s contained in the parameter space, denoted $\bar{\omega}(t)$, in which each possible *tfd* is weighted by the sum of all joint posterior probabilities apportioned to it.

     At first glance, this approach to posterior inference may seem analytically tedious if not entirely intractable, given (a) the multidimensionality of the problem, (b) the irregularity of the shape of the joint posterior distribution over the parameter space, and (c) the many-to-one relationship between parameter space locations and *tfd*s. In such seemingly problematic contexts,

plug-in parameter estimates derived through Monte Carlo (MC) simulation is often the favored shortcut (Hoff, 2009: 53; Gelman et al., 2013: 262). In the context of sample *tfd* estimation, such a simulation would proceed as follows:

1. Per iteration
    a. Make a single guess (denoted by the superscripted index $g$) at the sample's timestamps by sampling from the joint posterior probability distribution:

    $$\{\boldsymbol{\tau} = \mathbf{t}\}^{(g)} \sim p(\boldsymbol{\tau}|\mathbf{r}, \mathbf{s}, \mathbf{m}) \quad (13),$$

    where $\{\boldsymbol{\tau} = \mathbf{t}\}^{(g)}$ abbreviates a vector of $n$ guessed-at timestamps, i.e. one out of all possible locations in the parameter space per Eq. 6, and $p(\boldsymbol{\tau}|\mathbf{r}, \mathbf{s}, \mathbf{m})$ abbreviates the full joint posterior distribution.
    b. Summarize the temporal distribution for the $g$th simulated sample by calculating a mixture distribution comprising $n$ degenerate probability distributions:

    $$\omega^{(g)}(t) = \frac{\sum_{i=1}^{n} \delta\left(t|\tau_i^{(g)}\right)}{\int_{-\infty}^{\infty} \sum_{i=1}^{n} \delta\left(u|\tau_i^{(g)}\right) du} = \frac{1}{n} \times \sum_{i=1}^{n} \delta\left(t|\tau_i^{(g)}\right) \quad (14),$$

    where $\delta\left(t|\tau_i^{(g)}\right)$ is the degenerate probability density function,

    $$\delta\left(t|\tau_i^{(g)}\right) = \begin{cases} \infty & t = \tau_i^{(g)} \\ 0 & t \neq \tau_i^{(g)} \end{cases} \quad (15),$$

    which for the sake of computational ease may be exchanged for the corresponding probability mass function

    $$\delta\left(t|\tau_i^{(g)}\right) \propto \begin{cases} 1 & t = \tau_i^{(g)} \\ 0 & t \neq \tau_i^{(g)} \end{cases} \quad (16)$$

2. Repeat the per-iteration procedure a large number of times, $G$.
3. Across all $G$ guesses, calculate the average temporal distribution, denoted $\widehat{\overline{\omega}}(t)$, this being the plug-in estimator of the model-averaged *tfd* function $\overline{\omega}(t)$:

$$\hat{\bar{\omega}}(t) = \frac{1}{G} \times \sum_{g=1}^{G} \omega^{(g)}(t) \quad (17)$$

The plug-in estimator $\hat{\bar{\omega}}(t)$ converges toward the true model average $\bar{\omega}(t)$ as $G$ increases per the law of large numbers (Christian and Casella, 2010), though the computation time required to complete such a simulation also increases. In practice, MC simulations may be attenuated once a sufficient size of $G$ has been reached to insure negligible change in the plug-in estimate as $G$ continues to increase. However, even the time necessary to reach such an attenuation threshold is unnecessary, because as it turns out, *spd*s calculated per Eq. 5 identify with analytical exactitude the limiting case of this MC simulation as $G$ approaches infinity, assuming that Eq. 5 has been normalized:

$$\omega_{spd}(t) = \lim_{G \to \infty^-} \hat{\bar{\omega}}(t) \quad (18).$$

Fig. 4 illustrates the effect of simulation size on the plug-in estimate of the model average as $G$ increases across four orders of magnitude. The sample under consideration is the reduced IKIP-KBP sample for the Kuril Archipelago presented by Fitzhugh et al. (2016). As anticipated by Eq. 18, the *spd* calculated for this data set by Fitzhugh et al. (2016) appears to anticipate the eventual outcome of the plug-in estimate as the MC simulation increases in size and time.

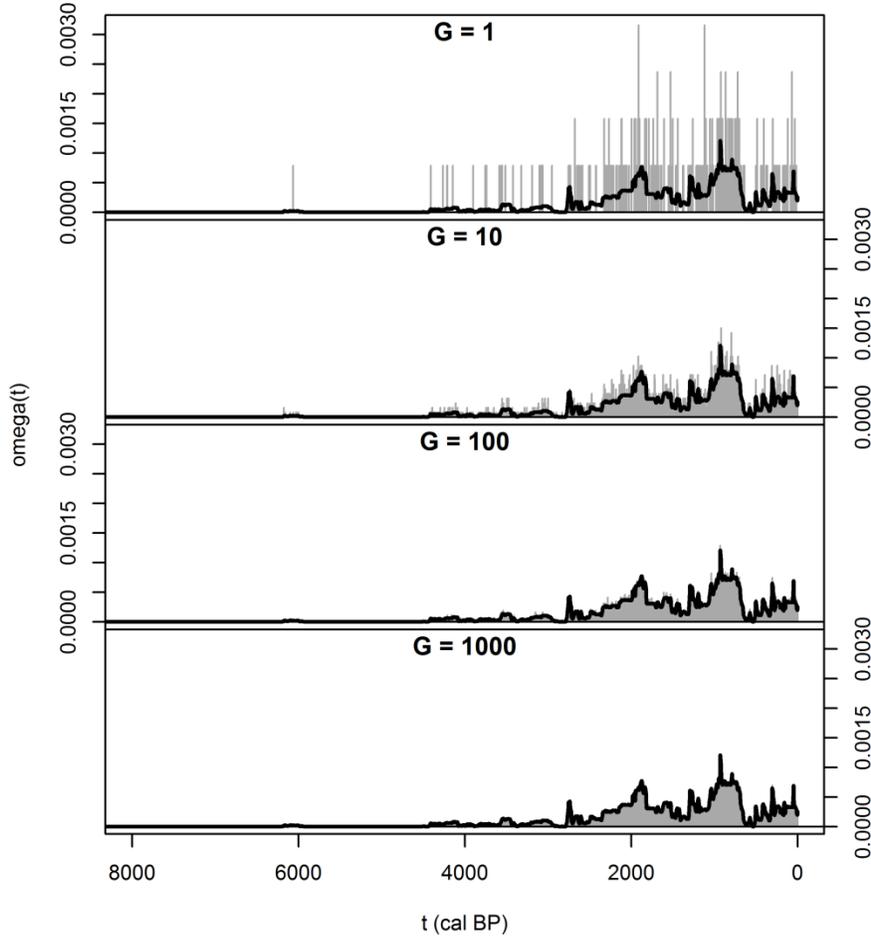

**Fig. 4.** Four plug-in estimates of $\widehat{\overline{\omega}}(t)$ based on MC simulations of increasing size ($G$ = 1, 10, 100, and 1000, respectively), calculated per Eqs. 13-14 and 16-17. The dataset comprises 253 individual and pooled $^{14}$C ages from the Kuril Archipelago, located between Hokkaido, Japan and the Kamchatka Peninsula, Russia, as reported by Fitzhugh et al. (2016). The *spd* derived from the same dataset is shown in black to illustrate the convergence of the plug-in estimates toward this distribution as simulation size increases.

In short, the *spd* is best interpreted as a "degraded" sample distribution. Such degradation is achieved through model-averaging, which provides a much more intuitively tractable alternative to the credible-region approach, which proves exceedingly problematic in the characteristically high-dimensional context of dTFA. While all possible *tfd*s are folded into the *spd*, the more probable of these exert a greater influence on the morphology of the resulting *tfd* than do the less probable, per the joint posterior distribution over the timestamp parameter space.

## 3. Random sampling error and kernel density estimation
### 3.1. Motivation

In contrast to probability summation, kernel density estimation was developed to mitigate uncertainty arising from random sampling error. It should also be noted that kernel density estimation arose not in dTFA but rather in general statistics, as an alternative to parametric model-fitting (cf. Silverman, 1986; Sheather and Jones, 1991; Wand and Jones, 1995; Jones et al., 1996; Sheather, 2004; Scott, 2015). To understand why an alternative approach to parametric model-fitting was deemed desirable enough to warrant the development of kernel density estimation, a more detailed review of the parametric approach is warranted.

At the most general level, the definitive operation of probability density estimation is the pursuit of estimates of the probability functions underlying particular data sets across their sample spaces. In the generic univariate case, such probability density estimates are denoted $\hat{f}(x)$, or $\hat{f}(t)$ in the case of dTFA. Given the continuity of the domain of $\hat{f}(x)$, this task is theoretically infinite in scope, necessitating the acceptance of some strategy to reduce this scope to something tractable. Here, the strategic virtue of parametric model-fitting is that it recasts the problem of infinite probability density estimation as a finite parameter estimation problem, founded on two key assumptions:

1. The probability distribution underlying a given dataset conforms closely to a relatively simple (usually closed-form) mathematical expression.
2. The case-specific probability distribution embodied by this expression may be rendered more flexible through generalization, i.e. by allowing certain of the equation's terms, factors, bases, exponents, etc. to vary between different dataset contexts (becoming constant only when a particular dataset is analyzed). These generalized terms, factors, etc. are the model's parameters, which may be generically denoted $\theta$ for the single-parameter model or abbreviate $\boldsymbol{\theta}$ for the multi-parameter case (i.e. where $\boldsymbol{\theta} = \{\theta_1, \dots, \theta_k\}$).

When specific parameter values are known for a given model and data generating context, this knowledge sufficiently implies the full series of densities or masses constituting the probability distribution. In such cases, parametrically conformable probability functions may be denoted as conditional probabilities $f(x|\boldsymbol{\theta})$. Conversely, when the parameters are unknown, the model may be fitted to a given dataset by searching across the parameter space for the combination of parameters that best accounts for the data. Different approaches for making such estimates (denoted $\hat{\boldsymbol{\theta}}$) are available depending on what the analyst regards to be a valid measure of a best fit, but the point is that reporting the finite set $\hat{\boldsymbol{\theta}}$ provides a concise shorthand alternative for the otherwise impossible task of reporting the infinite set of probability densities implied by these parameters, $f(x|\hat{\boldsymbol{\theta}})$. Parametric models also afford the ability to calculate likelihood functions, allowing the analyst to marshal probability-theoretic first principles in the service of parameter estimation. Finally, mechanistically meaningful interpretations can often be assigned to the fitted model parameters themselves, in certain research contexts.

However, in many research contexts, investigators either lack *a priori* reason to privilege one parametric model over others or may even have reason to believe that the data generating process (DGP) underlying their data is so irregular that it defies the rigidity even of the most flexible parametric models (e.g., in the case of multimodal distributions). In such cases, a more flexible alternative to the parametric approach is thus desirable.

It is largely in response to this demand that the desire for optimized histograms arose – that is, histograms whose origins and bin widths are algorithmically selected to provide robust if granular estimates of the underlying probability distribution (Freedman and Diaconis, 1981; Silverman, 1986: 7-11; Scott, 2015: 51-99). Even so, the granularity of histogram-based estimates often limits their utility in research contexts requiring high-resolution scales of analysis. Kernel density estimation was thus developed as a fine-grained refinement of and alternative to histogram-based estimation, i.e. by substituting the histogram's discrete-interval probability estimate for the *KDE*'s exact-point estimate (Silverman, 1986: 7-19).

In dTFA, parameteric population growth models have long enjoyed currency, most notably the exponential and logistic growth models (see discussion in Brown, 2017: 99-100). Over time, however, the use of such models has transitioned from the status of elegant and ecologically interpretable models to foils; even when the best-fitting parameters for such models are identified for given datasets, model evaluation criteria still indicate marked disparities between model prediction and observed data, suggesting that such models fail to provide particularly realistic approximations of the underlying DGP, parsimonious as they may be. It is perhaps largely due to this development that kernel density estimation has emerged into dTFA (Louderback et al., 2010; Grove, 2011; Tallavaara et al., 2014; Tallavaara, 2015; Fitzhugh et al., 2016; Weitzel and Codding, 2016; Brown, 2017; Baxter and Cool, 2016. For earlier applications in archaeology more generally, see Baxter et al., 1997; Baxter and Cool, 2010).

*3.2. Constituent distribution*

The constituent distribution underlying kernel density estimation is the 'kernel.' By convention, the kernel comprises a continuous time series of kernel functions, $K(t|\tau_i, h)$, which integrate to unity across the timeline, typically exhibiting a symmetrical and monotonically non-increasing shape around the kernel center. The particular form of the kernel used in any particular application is a matter of choice, having little to do with the unknown shape of the probability distribution underlying the sample. Choice of kernel shape seemingly exerts a negligible influence on the success of the probability density estimates produced by this method, at least when evaluated in terms of mean integrated square errors (MISEs; Silverman, 1986: 42-43). Consequently, a single kernel shape is customarily used for all *n* constituent kernels in the sample as a matter of computational efficiency and consistency. Commonly used forms include the Gaussian or normal, Epanechnikov, rectangular or uniform, triangular, and Laplace or double-exponential kernel functions.

The *i*th kernel takes the known or otherwise stipulated timestamp for the *i*th data point in the sample ($\tau_i$) as a location parameter, anchoring the center of the kernel to the timeline. Conversely, the standard approach assigns a single scale or smoothing parameter – labeled 'bandwidth' and denoted $h$ – to all *n* kernels constituting the sample (but see Silverman [1986: 21-23, 100-110] for alternative, kernel-by-kernel bandwidths). Selection of this bandwidth is typically data-driven, accomplished through the application of one or another algorithm. While several alternative algorithms are available, each with its own merits and limitations, all tend to respond to smaller sample sizes and/or more diffuse samples by selecting larger bandwidths. Ideally, such algorithms are not only data-driven but adaptive, i.e. optimizing the balance between 'under-smoothing' (failing to remove artificial structures from the estimated probability distribution) and 'over-smoothing' (removing real structures from the estimate).

Note that the fixed kernel shape and bandwidth between all $n$ kernels differs from the behavior of the *spd*'s constituent probability distributions, whose shapes and/or scales vary almost without exception between data points (excepting those with identical lab measures and errors, e.g. $r_i$ and $s_i$ in the case of $^{14}$C dating). Likewise, the role of the *i*th timestamp is markedly different between the two approaches: while in the case of probability summation the *i*th timestamp is the *estimand* (= unknown value under estimation) for the *i*th (posterior) probability distribution, in the case of kernel density estimation it instead serves as a known or otherwise stipulated location parameter for the *i*th kernel.

*3.3.* tfd *nomenclature and generative protocol*

Temporal distributions resulting from the aggregation of kernels are known as kernel density estimates (*KDE*s), whose definitive temporal density function is denoted $\hat{f}(t)$, calculated

$$\hat{f}(t) = \frac{1}{n} \times \sum_{i=1}^{n} K(t|\tau_i, h) \quad (19).$$

Fig. 1 might be viewed as an example of a *KDE* comprising two Gaussian kernels, save that bandwidth varies between the two kernels. Eq. 19 allocates equal fractions of the *tfd*'s total mass between all $n$ constituent distributions, with the *i*th fraction of mass diffusing symmetrically outward from the kernel's center in a distance-decaying manner, this center being dictated by the timestamp attributed to the *i*th observation in the sample. The normalizing constant in Eq. 19 (i.e., the reciprocal of sample size) insures that the estimated density function $\hat{f}(t)$ integrates to unity, in line with proper probability density functions. Graphically speaking, summing across a set of kernels positioned at the sample timestamps has three effects deemed desirable from the standpoint of probability density estimation: it fills in the gaps separating observed timestamps; it shows modes around those intervals of the timeline exhibiting the most densely clustered concentrations of observed timestamps; and it produces density estimates showing a high autocorrelation for small lags, diminishing as lag increases. Archaeological examples of *KDE*s generated following standard protocols are provided by Louderback et al. (2010), Tallavaara et al. (2014; Tallavaara, 2015), and Weitzel and Codding (2016).

*3.4. Interpretation*

The fraction of mass allocated to each constituent kernel is interpreted as a fraction of the probability distribution's total mass. As noted above, the position of each of these fractional masses over the timeline is data-driven, guided by the temporal distribution of the sample in a rule-of-thumb framework. Specifically, the standard approach to kernel density estimation involves the following rules, none of which are immutable (none are rooted in the first principles of probability theory):

(1) The total probability mass of the distribution is evenly apportioned among all $n$ constituent kernels, giving each a mass equaling unity (before normalization) or the reciprocal of $n$ (resulting in normalization);
(2) Each constituent kernel is anchored on a single data point in the sample, and each data point in the sample is assigned a single kernel;
(3) A single, arbitrarily selected kernel shape is assigned to all $n$ observations in the sample;
(4) A single scale parameter is assigned to all kernels, selected through the application of an algorithm sensitive to various attributes of the sample distribution (e.g., sample size, range, roughness, etc.).

While these rules of thumb are arbitrary, they have continuing currency because their ability to produce robust density estimates across a wide variety of sampling scenarios has been repeatedly demonstrated, mostly in the context of simulation experiments. Consequently, kernel density estimation continues to be favored over parametric model-fitting in cases where parametric conformability is in doubt. In light of the unremarkable performance of parametric models in dTFA, it is likely that kernel density estimation will continue to gain traction here, alongside dynamic growth models (Brown, 2017).

**4. Two protocols for jointly addressing chronometric and random sampling error**

The operation both of chronometric and random sampling error in the sample *tfd*'s DGP necessitates a propagation-of-uncertainty perspective (Marzouk and Willcox, 2015) on their mitigation, in which separate measures are required to address each problem in its own right. While the compounded application of such measures will inevitably produce *tfd*s that are more diffuse than either the *spd* or the *KDE*, such dispersion is a necessary evil of any inferentially robust operation in dTFA. Importantly, the (imperfect) algebraic resemblance holding between probability summation and kernel density estimation should not be mistaken to mean that the application of one doubles as the application of the other.

The main challenge facing archaeologists and paleontologists in attempting to apply kernel density estimation to their *tfd*s is the unmet need for exactly expressed timestamps. Informally, this may be accomplished following the approach described by Hommon: "pin [the *spd*] to the wall, walk to the other side of the room and look at it with eyes slightly out of focus. Characteristics of the curve that are still visible when we follow these directions should tell us something useful about the archaeological record and [the study region's] precensal tale. This is what I call the 'step back and squint' procedure, or 'squintosis', a technique of generalist intuitive statistics …" (Hommon, 1992: 152).

Of course, a formally rigorous approach will be more intellectually demanding than this approach. One recently popular, seemingly successful work-around for the unmet need for exact timestamps is the adoption of single-point estimates derived from probabilistically expressed timestamp estimates, e.g. Tallavaara's (2015: 26, Fig. 3) and Weitzel and Codding's (2016: 4) respective *KDE*s based on posterior medians of $^{14}$C dates. In theory, posterior means or MAPs may be used in the same capacity (cf. Tallavaara, 2015: 26).

However, the question arises as to whether single-point estimates based on probabilistic timestamps sufficiently fulfil the stipulated imperative to incorporate chronometric uncertainty into the *tfd*. Fig. 5 presents *KDE*s generated for the Kuril sample presented in Fig. 4, one each based on the posterior means, posterior medians, and MAPs derived from the sample's constituent posterior distributions. Each *KDE* is superimposed over the Kuril *spd* for comparison. Note that all three *KDE*s are smoother in shape than the *spd*, resulting from the relatively greater smoothness of the Gaussian kernel function used in these *KDE*s vis-à-vis the *spd*'s constituent posteriors. As a result, the sharpness of peak and trough structures in the *spd* are dampened to a degree in the *KDE*s. However, also observe that the *KDE*s maintain most of the main structures observed in the *spd*, and the temporal dispersion of the *KDE*s over the timeline is not noticeably more diffuse than that of the *spd* (cf. Louderback et al., 2010: 370). Choice of point estimate also leads to variability in the locations of several of the *KDE*s' respective high-resolution structures, as well as their magnitudes relative to one another, though bandwidths selected for all three based on the same algorithm are nearly identical.

It should be noted here that the bandwidth selection algorithms applied by Tallavaara (Tallavaara et al., 2014; Tallavaara, 2015: 26; algorithm not identified) and by Weitzel and Codding (2016: 4; Sheather-Jones method) lead to smoother *KDE*s than those presented in Fig. 5, which is based on the unbiased cross-validation algorithm (Scott, 2015), which has led Tallavaara to propose that kernel density estimation supersedes probability summation as a means of producing demographically reliable proxies (Tallavaara, 2015: 26).

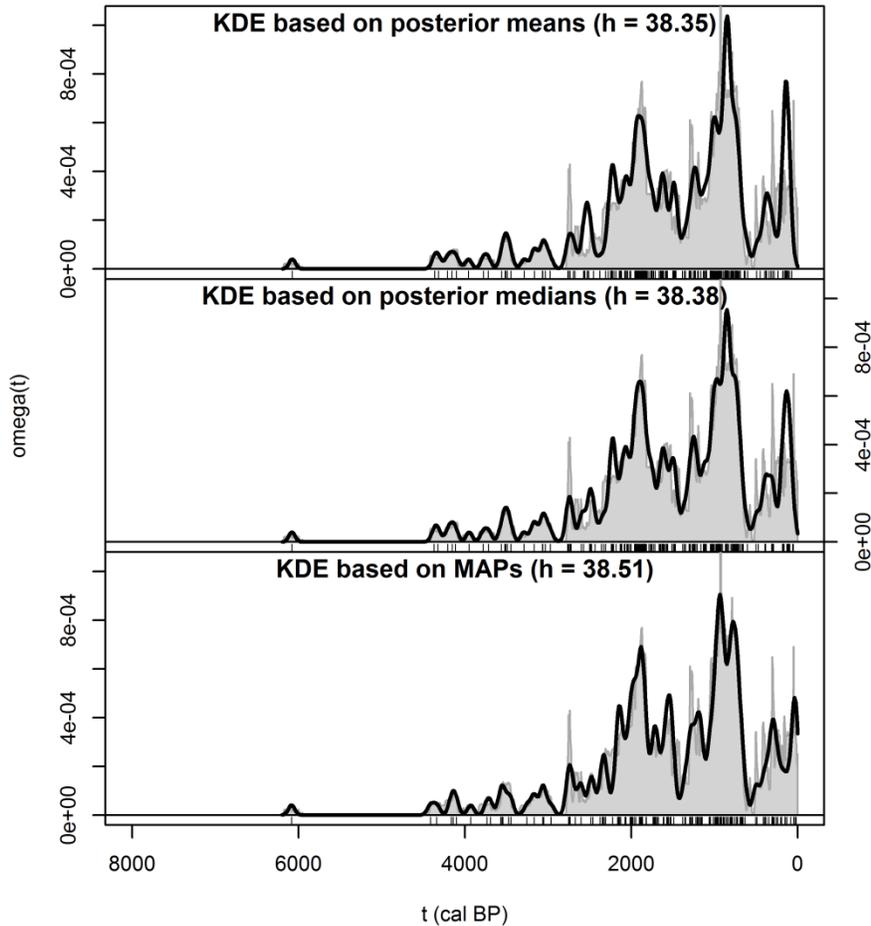

**Fig. 5.** Comparison of *KDE*s generated from the posterior means (upper panel), posterior medians (middle panel; cf. Tallavaara, 2015: 26, Fig. 3; Weitzel and Codding, 2016: 4), and MAPs. All *KDE*s aggregate Gaussian kernels, with the scale parameter selected by the unbiased cross-validation method (Scott, 2015). Mean, median, and MAP point estimates are shown by the rug plots under each panel.

Given the inconsistent behaviors exhibited between *KDE*s based on these different classes of posterior point estimate, the question arises as to which of them can most consistently lead to accurate probability density estimates, both within and between data sets. More importantly, we are reminded of a point already made over a decade ago by Telford et al. (2004): the choice to work with point estimates necessarily means the acceptance of incomplete summaries of the fruit of our timestamp estimation efforts, in effect sacrificing the containment of uncertainty that probabilistic estimates are intended to afford, and it cannot go unrecognized that Weitzel and Codding themselves concede that kernel density estimation "does not incorporate the complete calibrated distribution" (2016: 4); by extension we might remind ourselves that probability summation does not incorporate the kernel or any other means of recapturing data points missed through random sampling error. Thus, as already stated, if we wish to contain the compounding uncertainty entailed by the operation both of chronometric and random sampling error, a more successful solution to this challenge would require us to apply propagation-of-uncertainty protocols (Marzouk and Willcox, 2015). The present author has recently described and applied two alternative protocols toward such ends: respectively, weighted and composite kernel density estimation (Fitzhugh et al., 2016; Brown, 2017).

The composite kernel density estimation procedure described by Brown (2017) closely resembles the hypothetical Monte Carlo simulation described in Section 2 above. The crucial difference comes in how the per-guess sample distribution is summarized: instead of the mixture of degenerate distributions described in Eqs. 14-16, composite kernel density estimation favors the *KDE*, which is interpreted as a rule-of-thumb guess at the probability distribution underlying the $g$th guessed-at sample distribution, in line with Section 3:

$$\hat{f}^{(g)}(t) = \frac{1}{n} \times \sum_{i=1}^{n} K\left(t | \tau_i^{(g)}, h^{(g)}\right) \quad (20).$$

The final *tfd* is then generated by averaging across a large number of such *KDE*s, $G$:

$$\hat{f}_{composite}(t|\mathbf{r},\mathbf{s}) = \frac{1}{G} \times \sum_{g=1}^{G} \hat{f}^{(g)}(t|\mathbf{r},\mathbf{s}) = \frac{1}{G \times n} \times \sum_{g=1}^{G} \sum_{i=1}^{n} K\left(t|\tau_i^{(g)}, h^{(g)}\right) \quad (21)$$

where $\hat{f}_{composite}(t|\mathbf{r},\mathbf{s})$ is a plug-in estimator of the average of a theoretically infinite set of per-guess density estimation functions (compare Eq. 17). The distribution comprising the full series of such composite functions is labeled the composite kernel density estimate (*CKDE*). Note that, while the bandwidth $h^{(g)}$ is understood to hold constant between all $n$ kernels within the $g$th guess per Eq. 20, this parameter will virtually always differ between guesses as the distributional properties of each guess differ from those of the others (though typically not by very much). Because Eq. 20 replaces Eqs. 14-16 in this procedure, the *spd* no longer presents the limiting case of the plug-in estimator given in Eq. 21, and no other analytically tractable expression for this limiting case has yet been identified; at present it is only accessible through MC simulation. Also note that, while this procedure makes use of the same building blocks as does the *spd*, and while the resulting *CKDE* appears to smooth the often rough morphology of the *spd*, no *spd* is actually generated at any step of the protocol.

The upper panels of Fig. 6 illustrate to single-$g$ *KDE*s derived from the Kuril dataset. The lower panel of this figure illustrates 1000 such guesses, the MAP-based *KDE* (this being the *KDE* estimated for the single most likely guess across the timestamp space implied by the joint probability distribution per Eq. 12), and the *CKDE* determined by all 1000 simulation guesses (i.e., excluding the MAP-based *KDE*). Note that, unlike the three *KDE*s illustrated in Fig. 5, the dispersion of the *CKDE* is considerably more diffuse than that of the *spd* calculated for the sample, showing far fewer and more rounded peak and trough structures.

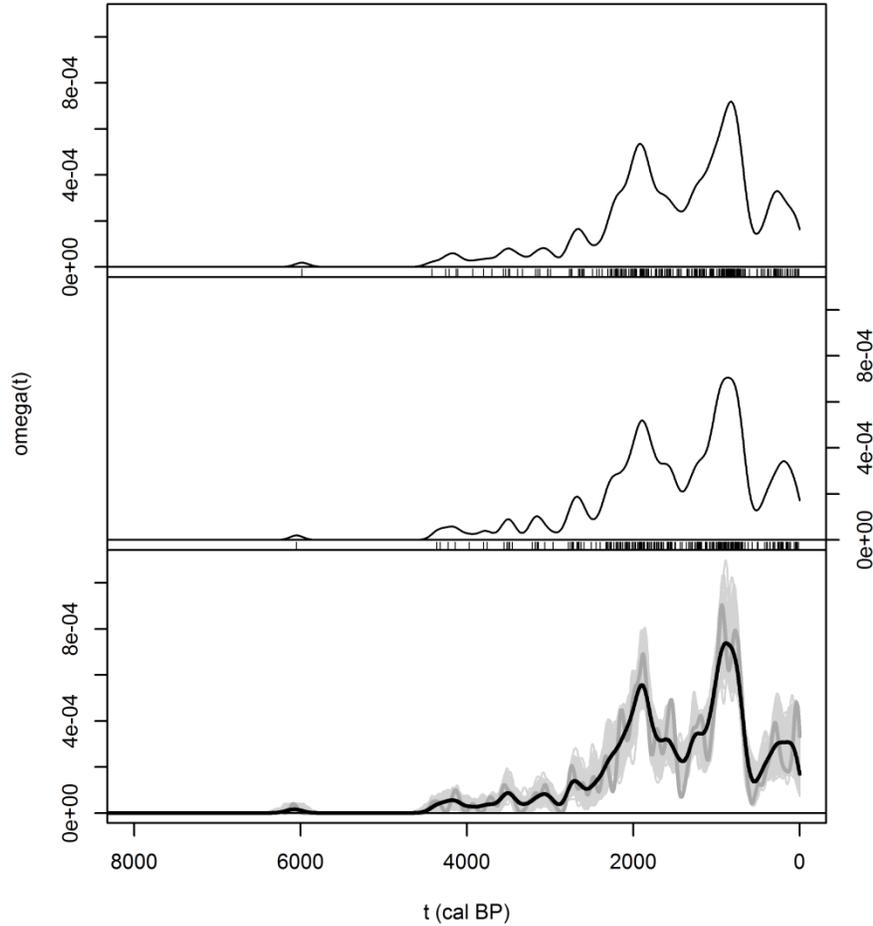

**Fig. 6.** Upper panel: *KDE* (black curve) based on a single MC guess at the true temporal distribution of the Kuril dataset (rug plot), assuming a Gaussian kernel and applying the unbiased covariance method for scale parameter selection. Middle panel: *KDE* (black curve) based on a second MC guess at the true temporal distribution of the Kuril dataset (rug plot), assuming a Gaussian kernel and applying the unbiased covariance method for scale parameter selection. Lower panel: 1000 *KDE*s (light gray curves) based on 1000 MC guesses at the true temporal distribution of the Kuril dataset (the guesses shown in the upper and middle panels included), all assuming a Gaussian kernel and applying the unbiased covariance method for scale parameter selection. The *KDE* based on the dataset's MAPs (shown in Fig. 5) is denoted by the dark gray curve, describing the single most likely guess out of all possible guesses. The *CKDE* (black curve) is the average of all 1000 individual *KDE*s. (Compare Crema, 2012: Fig. 6c-d; Baxter and Cool, 2016: Figs. 1, 3, and 4b).

Unlike the composite kernel density estimation approach, the approach nominally introduced by Fitzhugh et al. (2016: Supplementary Material) does involve the construction of a *spd* from the sample as an intermediate step. The final step then involves the calculation of a moving, negative distance-weighted average across this *spd*. The product of this protocol may be interpreted as a continuous, weighted *KDE*, in which a kernel is centered on every point along the timeline, each of whose mass is modulated according to the fraction of the sample allocated to that point by the *spd*:

$$\hat{f}_{weighted}(t|\mathbf{r},\mathbf{s}) = \frac{\int_{\infty}^{-\infty} \omega_{spd}(u)\, K(t|u,h)\, du}{\int_{\infty}^{-\infty} \omega_{spd}(u)\, du} \quad (22),$$

noting that the right side of the Eq. 22 simplifies to its numerator in the case that the mass of the *spd* is normalized to unity. Given the convention of presenting and storing *spd*s over a discretized timeline, this function is best approximated through summation (Fitzhugh et al., 2016: Supplementary Material, Eq. 3). Fitzhugh et al. (2016) used a Laplace or double-exponential kernel

$$K(t|u,h) = \frac{0.5}{h} \times e^{-\frac{|t-u|}{h}} \quad (23)$$

(noting that the negative sign was errantly omitted in the original presentation of this function) and applied an experimental data-driven bandwidth selection algorithm guided by sample size and one distributional property of the *spd*, the interquartile range (*IQR*):

$$h = -\frac{1}{\ln(0.05)} \times \mathrm{IQR}[spd] \times n^{-\frac{1}{6}} \quad (24).$$

This algorithm chooses a larger bandwidth either as the sample grows more diffuse (IQR increases) or sample size decreases, or both.

    Fig. 7 reproduces the weighted *KDE* presented by Fitzhugh et al. (2016) for the Kurils and compares it to the *CKDE* illustrated in Fig. 6, as well as the Kuril *spd* for comparison. Once again note that the weighted *KDE* is considerably more diffuse than the *spd*. It also shows fewer and less pronounced structures than does the *CKDE*, owing to a large degree to the disparate bandwidth selection algorithms underlying the two and potentially also the difference between the Gaussian and Laplace kernels used in the two. The relative efficiencies of the two approaches are a matter deserving greater attention in the future. On this point, note that the importance of per capita growth rates in demography warrants the measurement of MISEs not on the *KDE* in the raw but rather the log-transformed *KDE*.

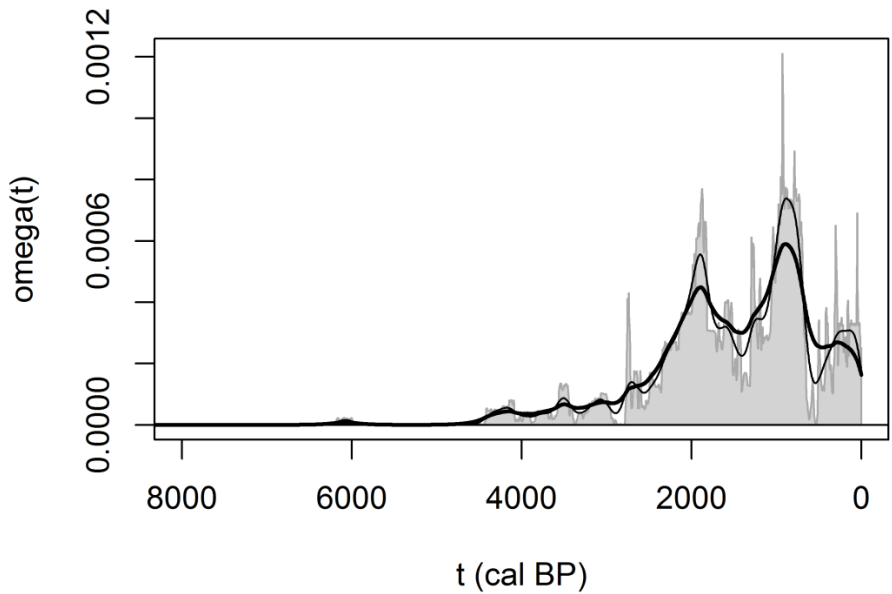

**Fig. 7.** Comparison of the *spd* (gray polygon), *CKDE* (thin gray curve), and weighted *KDE* (thick gray curve) generated from the Kuril dataset. The *CKDE* is the same as in Fig. 6, lower panel. The weighted *KDE* is calculated per Eqs. 22-24, reproducing the one presented by Fitzhugh et al. (2016: Fig. 4).

    The interpretation of smoothed *spd*s (= weighted *KDE*s) is as follows: recalling that the *spd* expresses a degraded sample distribution, in which a fraction of the sample size may in theory be attributed to any given point along the timeline, consideration is similarly warranted of a continuous sequence of kernels assigned to each and every point along the timeline. However, because the *spd* allots a greater share of the sample mass to some locations along the timeline than others, it is likewise appropriate to privilege the kernels located at certain locations over others with greater mass in the *KDE*, accomplished through the scaling of each kernel by the summed probability function at its center.

    Ultimately, while the relative merits and limitations of these two alternative approaches remain to be more fully explored in the future, their interpretation is much the same: both are regarded as good guesses to the shape of the probability distribution underlying the sample, guided by what we do know about the sample while also accommodating for what we don't.

## 5. Does $^{14}$C calibration introduce unacceptable structures into $^{14}$C-based *spd*s, and does *spd*-smoothing mitigate this purported problem?

    In recent years, considerable attention in dTFA has been paid to what may be called systematic calibration error in $^{14}$C-supported *spd*s. Specifically, given the nonlinear relationship between calendric and $^{14}$C years, the calibration of individual $^{14}$C age estimates in the form of posterior distributions over the calendric timeline and their subsequent aggregation into *spd*s

leads to the production of *tfd*s that are more likely to show dramatic peak and trough microstructures over certain intervals of the timeline than others (Kirch, 2007: 63-64; Shennan et al., 2013; Brown, 2015 and sources cited therein). From the standpoint of the *tfd*-based proxy census-taking that defines dTFA, such structures have been regarded as artificial and therefore undesirable, confounding the straightforward demographic interpretation of *spd*s.

The standing response to this seeming problem is the application of a moving average to the *spd*, resembling the weighted *KDE* protocol described above (Section 4, Eq. 22). Where calibration is believed to induce the anomalous re-sorting or distortion of mass under the *tfd* (a proposition that has been with us since the late 1980s or early 1990s Weninger, 1986: 32-33; McFadgen, 1994), the application of these moving averages is interpreted as undoing such effect. Perhaps the earliest articulation of this measure is in Weninger's (1986: 32-33) description of "Gauss-spreading" – formally equivalent to a weighted *KDE* comprising Gaussian kernels. While this Gaussian smoothing algorithm has been questioned by Culleton (2008) and Steele (2010) as it is implemented in CalPal, Williams' (2012) recent advocacy of a 500- to 800-year moving average of the *spd*, echoed in Shennan and colleagues' (2013) and Kelly and colleagues' (2014) 200-year moving averages, represents a revitalization of this measure, this time involving rectangular distance-weighting functions. The discord that has thus emerged regarding best practice (ideal kernel shape and bandwidth) in producing smoothed *spd*s thus warrants closer scrutiny; what, exactly, is actually wrong about the calibration-induced structures present in $^{14}$C-based *spd*s, and what measures are sufficient to fix them?

On this point, it bears repeating that the calibration of $^{14}$C age estimates in the framework of inverse uncertainty quantification (Marzouk and Willcox, 2015; Santosa and Symes, 2015) is founded upon, and therefore derives much of its inferential force from, the first principles of Bayesian probability theory. Consequently, the only elements of the inferential operation that are open to scrutiny in this framework are the selection (a) of a suitable forward map model (= calibration curve) to inform the likelihood function required by Bayes' theorem, and (b) of an appropriate prior distribution over the timeline, for each data point in the sample. If the potential critic has accepted these two crucial stipulations, little room then remains to bristle over the characteristics of the posterior distribution induced by the information brought into the inference; no matter how irregular the shape may be of the posterior distribution, such is the nature of the informed timestamp inference, as is true for the Bayesian solution of any calibration or inverse problem: it is a consequence of our best reasoning, not a symptom of failed reasoning and cannot therefore be regarded as unacceptable. By implication, any consequent measure taken to erase these structures must be regarded as counterproductive.

By extension, if we properly interpret the *spd* as and nothing other than a sample distribution, degraded as it may be, it is then exceedingly difficult to argue that anything has actually gone amiss in the process of *tfd* aggregation. Instead, the extreme, calibration-induced structures observed in *spd*s only become problematic when we push *spd*s to double as probability distribution estimates – a task for which sample distributions in general are poorly suited unless describing very large samples.

Instead, if the smoothed *spd* has any inferential merit, it is for the reasons discussed in Section 4 above, i.e. because it combines measures for mitigating chronometric and random sampling error. By extension, this procedure would be no less well applied to *spd*s based on other probabilistic timestamp estimates, e.g. luminescence dating, regardless of the fact that calibration-induced distributional structures are not salient features here.

Reframed in this way, it thus becomes clear that the disagreement regarding ideal kernel shape and bandwidth – Gaussian versus rectangular spreading, 800 versus 500 versus 200-year spreading – will be best resolved by evaluating these against one another in much the same way that other kernel density estimation techniques have been evaluated. By the same token, we are reminded that the passive reliance on *spd*s as robust estimates of their underlying probability distributions is a restrictive choice, obliging us to accept the stringent condition that such robustness will obtain only in the possession of very large samples. In the presence of adaptively data-driven CKDEs or weighted *KDE*s (= smoothed *spd*s), we are no longer constrained by the mandate that "summed probability plots based on less than 200-500 radiocarbon dates should be treated as provisional" (Williams, 2012: 581), though even in the case of the *CKDE* or weighted *KDE*, Williams's further statement that the *tfd* is "likely to change appreciably once larger datasets become available" (Williams, 2012: 581) still holds, insofar as the data-driven kernel smoothing parameter should decrease, resulting in the retention of more of the sample distribution's microstructures.

## 6. Uncertain site occupation durations and the calculation of the human occupation index

While Rick (1987) advocated the quantum of datable organic material (primarily charcoal) as an ideal unit of observation for dTFA, in practice archaeologists have instead continued to favor the occupied site as a more demographically reliable proxy (contra Attenbrow and Hiscock, 2015: 30; Drennan et al., 2015: 12-14). That is to say, the target event under quantification in the *tfd* is neither the acquisition, nor the use, nor the discard of dated material per se but instead the presence of one or more humans at a bounded location in space (a site) where such datable materials have accumulated through human agency. Under certain well-controlled conditions, the age estimated for such material may be convincingly used to anchor the site occupation to the timeline (Dean, 1978; Shott, 1992; Batt and Pollard, 1996; Pettitt et al., 2003; Kennett et al., 2008; Rieth and Hunt, 2008).

As challenging as the effort may be to convincingly demonstrate such chronometric equivalences, this challenge may be further amplified by the fact that any such individual timestamp can anchor human presence at a given site only to a singe point in the timeline, while site occupation is without exception an activity with duration. Note that this is not necessarily a challenge to dTFA focusing on the temporal enumeration of site occupations, since even an instant of human presence can be enumerated. Nevertheless, a more ambitious variant of site occupation-focused dTFA may wish to quantify the sites represented in a given data set for every point in time during which they were occupied.

Individual timestamp estimates for human presence do not provide sufficient leverage for such enumeration, at least not without the stipulation of further working assumptions to extend their instructiveness. That is to say, while we may desire knowledge of the opening and closing boundaries defining a given ($i$th) site occupation interval, $[t_{o,i}, t_{c,i})$ (where $t_{o,i}$ and $t_{c,i}$ denote the opening and closing temporal boundaries of the $i$th occupation interval), the timestamp provides information only on a single instant of occupation out of that interval, $\tau_i \in [t_{o,i}, t_{c,i})$, this being further blurred to the degree that we are uncertain of the true value of $\tau_i$; the timestamp does not in and of itself provide further information on the distance in time either between $t_{o,i}$ and $\tau_i$, or between $\tau_i$ and $t_{c,i}$.

The strategy proposed by Maschner et al. (2009) to mitigate such uncertainty – their "human occupation index" (HOI) method – resembles the MC-based composite kernel density estimation protocol discussed in Section 4, with three important differences:

(1) Rather than Eq. 20's kernel, the constituent distribution under summation for the $g$th guess is an "occupation window" (Maschner et al., 2009: 686), comprising a bounded, continuous series of scalable presence/absence indicator functions, denoted $o\left(t|\tau_i^{(g)}, h\right)$, where the $i$th guessed-at timestamp serves as a location parameter anchoring the center of the window to the timeline and $h$ denotes a scale parameter which in this case is equal to half the duration of the window;

(2) The mass apportioned to each occupation distribution is allowed to differ from those of the others:

$$o\left(t|\tau_i^{(g)}, h\right) = \Phi(a_i) \times I_o\left(t|\tau_i^{(g)}, h\right) \quad (25),$$

where $I_o\left(t|\tau_i^{(g)}, h\right)$ denotes an indicator function equaling 1 if $t$ falls within the $i$th occupation window

$$I_o\left(t|\tau_i^{(g)}, h\right) = \begin{cases} 1 & t \in \left[\tau_i^{(g)} + h, \tau_i^{(g)} - h\right) \\ 0 & t \notin \left[\tau_i^{(g)} + h, \tau_i^{(g)} - h\right) \end{cases} \quad (26)$$

and $\Phi(a_i)$ denotes the "conversion factor" (Maschner et al., 2009: 686), a linear transformation of site area (in m²) attributed to the $i$th occupation window –

$$\Phi(a_i) = \frac{a_i}{100} \quad (27)$$

– which is intended to capture inter-site population size differences under the assumption that site area serves as a linear proxy of this demographic estimand (Maschner et al., 2009: 686; but see Chamberlain, 2006: 127-128 for a discussion of nonlinear relationships between site population and area). By implication, the mass assigned to the $i$th occupation window is

$$\Phi(a_i) \times 2h = \int_{\infty}^{-\infty} o\left(t|\tau_i^{(g)}, h\right) dt \quad (28).$$

However, since $\Phi(a_i)$ is proportional to $a_i$ and all occupation windows share the same scale $h$, the ratio of masses between any two occupation windows is determined solely as the ratio of their respective site areas,

$$\frac{\Phi(a_1) \times 2h}{\Phi(a_2) \times 2h} = \frac{\frac{a_1}{100} \times 2h}{\frac{a_2}{100} \times 2h} = \frac{a_1}{a_2} \quad (29),$$

so that $a_i$ may be treated as the conversion factor with no actual need for the calculation of $\Phi(a_i)$, particularly given the virtue of dimensionlessness favored by Maschner et al. (2009: 686). By extension, $o\left(t|\tau_i^{(g)}, h\right)$ may be revised given the proportionality this implies:

$$o\left(t|\tau_i^{(g)}, h\right) \propto \begin{cases} a_i & t \in \left[\tau_i^{(g)} + h, \tau_i^{(g)} - h\right] \\ 0 & t \notin \left[\tau_i^{(g)} + h, \tau_i^{(g)} - h\right] \end{cases} \quad (30)$$

(3) $h$ is fixed at 50 across all $n$ occupation windows in the sample and all $G$ guesses in the simulation. While the implicit symmetry of mass projected in this way constitutes an arbitrary rule of thumb, it may be regarded as a reasonable choice balancing between those occupation windows where $\tau_i^{(g)}$ lies closer to $t_{o,i}$ and those where it lies closer to $t_{c,i}$. The 100-year interval implied by $h = 50$,

$$\left[t_{o,i}^{(g)}, t_{c,i}^{(g)}\right] = \left[\tau_i^{(g)} + 50, \tau_i^{(g)} - 50\right] \quad (31).$$

The validity of this decision is conditional on the assumption that this 100-year window is ethnographically credible, especially when projected backward over extensive tracts of time. Maschner et al. (2009: 686) note that modulation of $h$ results in greater or less overlap between occupation windows and smoother or rougher *HOID*s. However, since the goal of this protocol is to mitigate uncertainty surrounding the duration of occupation intervals rather than sampling error (as in kernel density estimation), modulation of overlap and smoothing should be seen as incidental features of the HOI approach.

Aggregation of the *tfd* through summation across all $n$ occupation windows, resulting in what may be labeled a human occupation index distribution (*HOID*), follows the standard summation equation (Eq. 3) –

$$\omega_{HOI}^{(g)}(t) = \sum_{i=1}^{n} o\left(t|\tau_i^{(g)}, h\right) \quad (32)$$

– noting that in this case the standard equation's scaling constant is set to 1 and thus simplifies out of the equation. Maschner and colleagues' MC simulation (labeled a "probabilistic resampling" simulation; 2009: 686)[3] takes $G = 1000$ iterations, and the average across all iterations results in a composite human occupation index distribution (*CHOID*):

$$\omega_{CHOI}(t) = \frac{1}{G} \times \sum_{g=1}^{G} \omega_{HOI}^{(g)}(t) = \frac{1}{G} \times \sum_{g=1}^{G} \sum_{i=1}^{n} o\left(t|\tau_i^{(g)}, h\right) \quad (33)$$

(compare Eq. 21). See Maschner et al. (2009: Fig. 8) for a visual representation of a *CHOID* generated from data for Sanak Island off the southern coast of the Alaska Peninsula.

## 7. Discussion and conclusion

The survey of protocols for aggregating *tfd*s presented here is offered for two reasons: first as a systematic overview and description of these prevailing and emerging dTFA methods, and second as a critical reminder that formal theories – including those supporting programmatic methods – are incomplete if unaccompanied by interpretation. While the formal resemblances between the five protocols reviewed here are unmistakable – strikingly so – each nevertheless exhibits detailed nuances separating it from the others, following from fundamental differences between their respective motivations and interpretations. It is thus hardly pedantic to attend to the need for their interpretive and operational separation.

---

[3] The likeness assumed here between the composite kernel density estimation approach and Maschner and colleagues' HOI protocol is based on the choice to interpret the expression "probabilistic resampling" (Maschner et al., 2009: 686) to mean 'Monte Carlo simulation' rather than 'resampling' in the strict sense (Good, 2006). While these two modes of simulation may be computationally similar to one another, their motivations and underlying assumptions differ markedly (Roberts and Casella, 2004; Good, 2006; Rubinstein and Kroese, 2008; Thomopoulos, 2013). In the context of the passage where this expression appears, such an interpretation is well-warranted, as unceremonious/uncharitable as it may at first appear, insofar as Maschner and colleagues intend this element of their approach to mitigate the problem of chronometric uncertainty, not sampling error toward which resampling methods are typically applied: "Given the fact that the estimated error associated with any given carbon date may well be plus or minus 100 years or so, it is possible to construct multiple estimates of population density, based on the estimated error of all the carbon dates. One method for dealing with this problem is to use the estimated error associated with each carbon date to define a possible window of occupation for each site based upon probabilistic resampling" (Maschner et al., 2009: 686). If this expression were instead interpreted in the strict sense, it is not clear which kind of resampling the authors may have in mind, though bootstrapping would be a reasonable guess (Efron, 1979), with some precedent in dTFA (Rick, 1987; Dye and Komori, 1992; Williams, 2012: 580; Zahid et al., 2016: 934), particularly given the quantile envelopes Maschner and colleagues (2009) present in their Fig. 8. However, such envelopes may just as easily be produced based on MC simulations (e.g., summarizing the spread of the simulated *KDE*s in the lower panel of Fig. 6 above; cf. Crema, 2012: Fig. 6c-d; Baxter and Cool, 2016: Figs. 1, 3, and 4b).

     Thus, to answer the question raised by Baxter and Cool (2016) of the reinvented wheel, it would be more accurate to say that it has been repurposed, to the advancement of dTFA's methodology. Or, if one prefers culinary figurative language, one might consider the egregious consequence of mistaking casseroles for cobblers; while preparation of both requires the application of recipes that are far more similar to each other than (for example) the procedure for making an ice cream sundae, it would nevertheless be an unfortunate mistake to assume that either the casserole or the cobbler can acceptably double for the other. Despite some similarities in technique between the two dishes, detailed differences in preparation and major differences in ingredients will result on one hand in a main course (in the case of the casserole) and a desert (in the case of the cobbler). To sufficiently round out the experience of the diner, the chef would be better-advised to take a multicourse approach than to attempt to convince their diner that the one is also the other on the basis of similar preparations alone.

     In most archaeological and paleontological case studies in dTFA, uncertainty surrounds both those data points included in the sample (in the form of chronometric uncertainty) and those that have been omitted from it (in the form of random sampling error). While *spd*s and *KDE*s respectively address these two problems individually, their joint mitigation requires separate operations dedicated to the containment of both sources of uncertainty. The *CKDE* and weighted *KDE* (= smoothed *spd*) approaches reviewed in Section 4 offer two such measures, both resulting in the aggregation of *tfd*s exhibiting markedly greater dispersion than do either *spd*s or *KDE*s alone (contrast Figs. 5 and 7).

     Conversely, the purported problem of calibration interference deserves no special effort at mitigation at all. On the contrary, if we concede that the peculiar shapes that characterize posterior distributions for calibrated $^{14}$C age estimates are induced by design – specifically that they are a consequence of the principled, informed updating of belief in a Bayesian framework – then little room remains to regard the sometimes dramatic oscillations characterizing *spd*s as artificial or unacceptable; they are there because they have to be, at least to the degree that we

a) trust the $^{14}$C lab measurement;
b) accept the forward map model(s) applied in the calibration;
c) accept the prior distribution(s) stipulated for the calendric ages of the data points in the sample; and
d) accept the interpretation of *spd*s as "degraded" sample distributions accommodating chronometric uncertainty and multidimensional timestamp estimation.

Instead, such oscillatory structures become problematic only in the case that we further push *spd*s to perform as probability distribution estimates. By implication, if the smoothing of *spd*s does have any uncertainty quantification merit, it is because it jointly addresses both chronometric and random sampling error – a fact that would be no less true of smoothed *spd*s aggregating other kinds of probabilistic age estimates than $^{14}C$ dates.

In contrast, the human occupation index approach recommended by Maschner and colleagues (2009) addresses yet another source of uncertainty, though one that is only relevant when the unit of observation under temporal enumeration is the occupied site. When the *tfd* is intended to quantify the number of occupied sites as this changes over time, possession of a single timestamp for a given site occupation interval does not sufficiently allow us to fully infer the duration or temporal location of such intervals. As a tentative patch for such uncertainty, Maschner and colleagues' (2009) recommended approach involves projecting occupation mass evenly and symmetrically backward and forward from the single timestamp, presumably up to an ethnographically reasonable length of time. The scaling of mass under each occupation window by site area is also recommended as a further means of bending site occupation *tfd*s to the service of demography. The main constraint of the HOI approach is the selection of the 100-year occupation window; this operation is an arbitrary choice which may work well in the context of kernel density estimation but is of questionable relevance as a patch for site occupation uncertainty. One alternative approach with some precedent is to subject each site represented in the sample to its own intra-site temporal frequency analysis prior to aggregation across the whole sample (Collard et al., 2010: 867; Shennan et al., 2013: 6; cf. Story and Valastro, 1977; Shott, 1992; Batt and Pollard, 1996; Hutchison and McMillan, 1997; Grier, 2006). A second approach is also conceivable, in which the number, temporal location(s), and duration(s) of site occupation intervals represented by the intra-site assemblage are estimated through the application of a latent, finite mixture modeling approach (McLachlan and Peel, 2000; Marin et al., 2005) in the framework of Bayesian parameter estimation and model selection (Bronk Ramsey, 2009a; Gelman et al., 2013: 165-195). This approach would be very similar to the fitting of noncontiguous-multiple-phase models described by Bronk Ramsey (2009a: 348), the key difference being the absence of phase assignment indices (such as *i* and *j* in Bronk Ramsey, 2009a: Eq. 25).

The identification of a *tfd* protocol that simultaneously addresses chronometric, random sampling, and site occupation uncertainty is a matter for future consideration. In theory, however, we should expect the output of such a protocol to be a more diffuse *tfd* still than either the *CKDE* (or weighted *KDE*) or the *CHOID*.

It should be noted that the formal resemblances discussed throughout this paper belong to methods hailing from an eclectic assortment of frameworks of statistical inference. As argued in Section 2, probability summation is fully interpretable in the framework of Bayesian probability theory, and as such most of its operations follow from probability-theoretic first principles. Conversely, the *KDE* belongs to nonparametric statistical inference, many of whose operations are based on flexible rules of thumb only (noting however that Bayesian statisticians are frequent consumers of kernel density estimation). The importance of resampling methods in dTFA – e.g. bootstrap estimates of variances and confidence intervals (Rick, 1987; Dye and Komori, 1992; Williams, 2012; Zahid et al., 2016), permutation tests (Crema et al., 2016), kernel bandwidth selection through cross-validation (Brown, 2017; cf. Scott, 2015) – brings another flavor of nonparametric statistics into the mix, whose coherence with or interpretability in terms of Bayesian principles remains to be sufficiently demonstrated (Gelman et al., 2013: 96-97). Pragmatically speaking, such eclecticism may not be a bad thing for dTFA, though theoretical purists may find occasion for greater self-examination here.

Further sources of uncertainty confronting dTFA research have been extensively outlined elsewhere, beginning with Rick's inaugural paper (Rick, 1987; Kirch, 2007: 63-64; Surovell and Brantingham, 2007: 1869; Shennan et al., 2013: 3; Kelly and Naudinot, 2014: 547; Brown, 2015: Table 2; Fitzhugh et al., 2016: 181-182). It is unlikely that the generic summation-of-constituent-distributions approach discussed throughout this paper can be convincingly applied to all of them, but it is a matter worth exploring.

*Acknowledgments*
The critical reflections presented in this paper germinated out of a handful of conversations with both Lisbeth Louderback and Marcos Llobera about kernel density estimation spread across several years (approximately 2007 through 2014), as well as ongoing conversations with Ben Fitzhugh about its application in our work in the North Pacific (since approximately 2010). Several of my colleagues at CSSCR have granted me their patient attention as I have struggled to better articulate the formal and conceptual dimensions of the arguments presented here. All errors (typos, errors in reasoning, misrepresentations of others' work) are mine alone.